\documentclass{article}
\usepackage[T1]{fontenc}
\usepackage{graphics}

\makeatletter

\providecommand{\LyX}{L\kern-.1667em\lower.25em\hbox{Y}\kern-.125emX\@}

\usepackage{graphicx}

\begin{document}

\title{The hot neutron star }

\author{Ilona Bednarek, Ryszard Ma\'{n}ka\\
 \textsl{Institute of Physics,}\\
 \textsl{University of Silesia,}\\
 \textsl{Uniwersytecka 4, 40-007 Katowice,}}

\maketitle
\begin{abstract}
In this paper the equation of state of hot neutron matter is calculated. Involving
the Oppenheimer-Volkov- Tolman equation global parameters of a neutron star
at the finite temperature are obtained. The objective of our work was to study
the influence of the temperature on the main parameters of a neutron star.
\end{abstract}

\section{Introduction}

This paper is concerned with a neutron star and its macroscopic parameters such
as the mass \( M \) and the radius \( R \) which are influenced by the temperature.
Both the mass and the radius as well as the cooling evolution are determined
first of all by the equation of state. \\
 Considering the matter of a neutron star one should take into account not only
neutrons but as the most elementary model has it neutrons, protons and leptons.
This paper presents a basic model of neutron star matter including interactions
among nucleons in the Hartree approximation \cite{walecka}\cite{weber}. Increasing
interest in neutron matter at finite temperature has been observed recently
in relation to the problems of hot neutron stars and of protoneutron stars and
their evolutions in particular. Theories concerning protoneutron stars are being
discussed in works by Prakash et. al. \cite{prakash}. Some other elements like
hyperons, mesons or quarks could be also found in the interior of a neutron
star but their relevance is not going to be included in our work. \\
 One can divide this paper into two parts. In the first one thermodynamic properties
of Fermi gas, which consists of neutrons, protons and leptons, are examined.
The properties of the quark matter at finite temperature were also looked into
in \cite{bm:1998}. The range of temperatures considered vary from 0-50 MeV.
Analytical forms for the pressure, energy density and fermion number density
has been calculated in order to solve the OTV equation. Solution of this equation
is the main subject of the second part.

\section{General Theory}

The aim of this paper is to present a rudimentary approach to the equation of
state for neutron stars at the finite temperature. In such an approach the neutron
star matter consists of electrically neutral plasma which comprises protons,
neutrons and electrons. The Lagrange density function in this model is given
by 
\begin{eqnarray}
{\mathcal{L}}=i\overline{\psi }\gamma ^{\mu }\partial _{\mu }\psi -\bar{\psi }M\psi +v<\bar{\psi }\psi >\bar{\psi }\psi -\frac{1}{2}v<\bar{\psi }\psi >^{2}+ &  & \label{lag} \\
i\sum ^{2}_{f=1}\overline{L_{f}}\gamma ^{\mu }\partial _{\mu }L_{f}-\sum ^{2}_{f=1}g_{f}(\overline{L}_{f}He_{Rf}+h.c.)-\frac{1}{2\kappa }R &  & \nonumber \label{lagg} 
\end{eqnarray}
 where \( \kappa =8\pi G/c^{4} \) and \( R \) is the Ricci curvature scalar.
The fermion fields are composed of neutrons, protons and electrons, muons and
neutrinos 
\begin{equation}
\psi =\left( \begin{array}{l}
\psi _{n}\\
\psi _{p}
\end{array}\right) ,\, \, \, L_{1}=\left[ \begin{array}{l}
\nu _{e}\\
e^{-}
\end{array}\right] _{L},\, \, \, L_{2}=\left[ \begin{array}{l}
\nu _{\mu }\\
\mu ^{-}
\end{array}\right] _{L},\, \, \, e_{Rf}=(e^{-}_{R},\, \mu ^{-}_{R})
\end{equation}
 and the Higgs field \( H \) takes the form of 
\begin{equation}
H=\frac{1}{\sqrt{2}}\left( \begin{array}{l}
0\\
V
\end{array}\right) 
\end{equation}
 Nucleon masses are given by 
\begin{equation}
M=\left( \begin{array}{cc}
m_{n} & 0\\
0 & m_{p}
\end{array}\right) 
\end{equation}
 The electron and muon masses equal \( m_{e}=g_{1}V/{\sqrt{2}},\, \, m_{\mu }=g_{2}V/{\sqrt{2}} \)
respectively where \( V=240 \) GeV. The model describes nuclear interaction
in the Hartree approximation for \( v\neq 0 \) \cite{walecka}. Introducing
the interaction one can obtain the effective nucleon mass 
\begin{equation}
\label{meff}
M^{*}=M-v<\bar{\psi }\psi >
\end{equation}
The constant term appearing in (\ref{lag})is responsible for the negative pressure
\begin{equation}
P=\frac{1}{2}v<\bar{\psi }\psi >^{2}
\end{equation}
This model is the simple approximation of the relativistic mean filed theory
\cite{walecka}. The equation (\ref{meff}) is a highly nonlinear equation due
to the average \( <\overline{\psi }\psi > \) which is defined as
\begin{eqnarray*}
<\overline{\psi }\psi >=\frac{M^{*}M^{2}}{\pi ^{2}}\int _{0}^{\infty }\frac{y^{2}dy}{\sqrt{y^{2}+\delta ^{2}}}\{\frac{1}{\exp (\beta (\epsilon _{k}-\mu ))+1}+ &  & \\
\frac{1}{\exp (\beta (\epsilon _{k}+\mu ))+1}\} &  & 
\end{eqnarray*}
The solution of the equation (\ref{meff}) help us to define the parameter \( \delta =m_{eff}/m \).
The term \( m_{eff} \) denotes fermion effective mass which for nucleons is
written as \( m_{eff}=M^{*} \)whereas for leptons \( m_{eff}=m \) thus \( \delta =1 \).
Fig. 1 shows the relation between the parameter \( \delta  \) and the dimensionless
Fermi momentum \( x \) for three different temperatures \( T=0,\, 10,\, 50 \)~
MeV. The simplest case corresponds to L2 parameter set \cite{walecka}. In the
relativistic mean filed theory it corresponds to \( m_{s}=520\, MeV \) and
\( g_{s}=10.47 \) what gives
\[
v=\frac{g_{s}}{m_{s}^{2}}\]
 
\begin{figure}
{\par\centering \resizebox*{15cm}{!}{\includegraphics{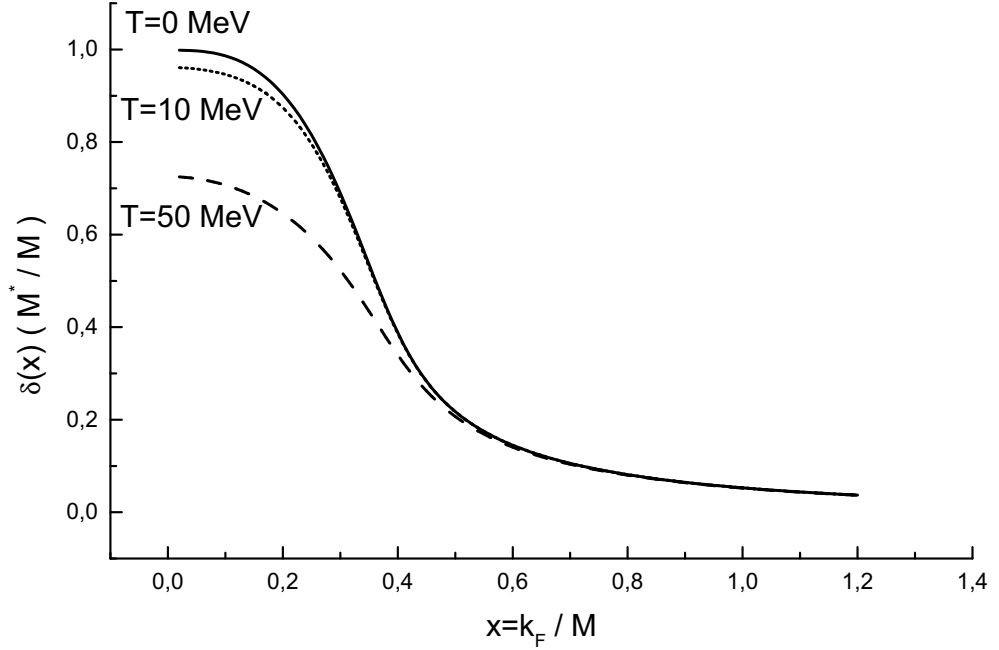}} \par}

\caption{The effective nuclear mass }
\end{figure}

The fermion number density, energy density and pressure inside the star are
defined locally as quantum averages. One can calculate such quantum averages
employing the following equation 
\begin{equation}
<A>=\frac{1}{Z}Tr(e^{-\beta ({\mathcal{H}}-\sum _{f}\mu _{f}N_{f})}A)
\end{equation}
 where \( A \) is an observable, \( {\mathcal{H}} \) is the Hamiltonian of
the system and 
\[
Z=Tr(e^{-\beta ({\mathcal{H}}-\sum _{f}\mu _{f}N_{f})})\]
 is a partition function. In this equation \( \beta =1/k_{B}T \). Let us now
define some operators indispensable in our calculations: the fermion number
operator 
\begin{equation}
N_{f}=\sum _{{\textbf {{{k}}}},\sigma }(c_{{\textbf {{{k}}}},\sigma ,f}^{+}c_{{\textbf {{{k}}}},\sigma ,f}-d_{{\textbf {{{k}}}},\sigma ,f}^{+}d_{{\textbf {{{k}}}},\sigma ,f})
\end{equation}
 where the index \( f \) stands for neutrons, protons and electrons (\( f=p,n,e \))
and \( c_{{\textbf {{k}}},\sigma ,f}^{+},c_{{\textbf {{k}}},\sigma ,f} \) and
\( d_{{\textbf {{k}}},\sigma ,f}^{+},d_{{\textbf {{k}}},\sigma ,f} \) are creation
and annihilation operators for particles and antiparticles respectively, the
fermionic Hamiltonian \( {\mathcal{H}} \) (ignoring the zero-point energy)
\begin{equation}
{\mathcal{H}}=\sum _{f}\sum _{{\textbf {{k}}},\sigma }\epsilon _{{\textbf {{k}}},f}(c_{{\textbf {{k}}},\sigma ,f}^{+}c_{{\textbf {{k}}},\sigma ,f}+d_{{\textbf {{k}}},\sigma ,f}^{+}d_{{\textbf {{k}}},\sigma ,f})
\end{equation}
 with \( \epsilon _{{\textbf {{k}}},f}=c\sqrt{\hbar ^{2}k^{2}+m_{f}^{2}c^{2}} \)
and the pressure with an isotropic distribution of momenta given by 
\begin{equation}
P_{f}=\frac{1}{3}\sum _{f}\sum _{{\textbf {{k}}},\sigma }\hbar kv_{{\textbf {{k}}},f}(c_{{\textbf {{k}}},\sigma ,f}^{+}c_{{\textbf {{k}}},\sigma ,f}+d_{{\textbf {{k}}},\sigma ,f}^{+}d_{{\textbf {{k}}},\sigma ,f})
\end{equation}
 where velocity equals \( v_{{\textbf {{k}}},f}=\hbar kc^{2}/\epsilon _{{\textbf {{k}}},f} \).
The mean number of fermions is determined by the following equation 
\begin{equation}
<c_{{\textbf {{k}}},\sigma ,f}^{+}c_{{\textbf {{k}}},\sigma ,f}>=\frac{1}{\exp (\beta (\epsilon _{{\textbf {{k}}},f}-\mu _{f}))+1}
\end{equation}
 where \( \mu _{f} \) stands for the fermion chemical potential. Neutrons,
protons and electrons are in \( \beta  \)-equilibrium which can be described
as a relation among their chemical potentials 
\begin{equation}
\label{eq3}
\mu _{p}+\mu _{e}=\mu _{n}
\end{equation}
 where \( \mu _{p} \), \( \mu _{n} \) and \( \mu _{e} \) stand for proton,
neutron and electron chemical potentials respectively. If the electron Fermi
energy is high enough (greater then the muon mass) in the neutron star matter
muons start to appear as a result of the following reaction
\begin{equation}
e^{-}\rightarrow \mu ^{-}+\nu _{e}+\overline{\nu _{\mu }}
\end{equation}
 The chemical equilibrium between muons and electrons can be described by the
condition
\begin{equation}
\mu _{\mu }=\mu _{e}
\end{equation}
 The equation (\ref{eq3}) together with the charge neutrality \( n_{e}+n_{\mu }=n_{p} \)
allows us to determine the equation of state in terms of only one parameter
\( x_{n} \). It defines the dimensionless Fermi momentum of a neutron \( x_{n}=(\hbar k_{n})/(m_{n}c) \).
Using the stated above conditions and the equation 
\begin{equation}
\label{eq10}
\mu _{f}=m_{f}c^{2}(1+x_{f}^{2})^{1/2}
\end{equation}
 the relations between \( x_{n}=x \) and \( x_{p} \) can be obtained \\
 In the macroscopic limit an integral is allowed to replace a sum 
\begin{equation}
\sum _{{\textbf {{k}}}}\rightarrow \frac{V}{(2\pi )^{3}}\int d^{3}k
\end{equation}
 and for each fermion the quantum averages \( n_{f} \) representing the particle
number density \( n_{f}=<N_{f}>/V \), \( \varepsilon _{f} \) the energy density
and \( P_{f} \) the pressure can now be written as 
\begin{equation}
n_{f}(\mu _{f},T)=n_{0}\Theta (r,y)
\end{equation}
\begin{equation}
\varepsilon _{f}(\mu _{f},T)=c^{2}\rho (\mu _{f},T)=\varepsilon _{0}\chi (r,y)
\end{equation}
\begin{equation}
P(\mu _{f},T)=P_{0}\Phi (r,y)
\end{equation}
 The forms of these quantum averages are determined by the functions \( \Theta (r,y) \),
\( \chi (r,y) \) and \( \Phi (r,y) \) which are presented below. 
\begin{eqnarray}
\Phi (r,y) & = & \frac{1}{3\pi ^{2}}\int ^{\infty }_{0}\frac{z^{4}dz}{\sqrt{z^{2}+\delta ^{2}y^{2}}}\{\frac{1}{\exp {(\sqrt{z^{2}+\delta ^{2}y^{2}}-ry)}+1}\\
 & + & \frac{1}{\exp {(\sqrt{z^{2}+\delta ^{2}y^{2}}+ry)}+1}\}\nonumber 
\end{eqnarray}
\begin{eqnarray}
\Theta (r,y) & = & \int ^{\infty }_{0}z^{2}dz\{\frac{1}{\exp {(\sqrt{z^{2}+\delta ^{2}y^{2}}-ry)}+1}\\
 & - & \frac{1}{\exp {(\sqrt{z^{2}+\delta ^{2}y^{2}}+ry)}+1}\}\nonumber 
\end{eqnarray}
\begin{eqnarray}
\chi (r,y) & = & \frac{1}{\pi ^{2}}\int ^{\infty }_{0}dz\sqrt{z^{2}+\delta ^{2}y^{2}}\{\frac{1}{\exp {(\sqrt{z^{2}+\delta ^{2}y^{2}}-ry)}+1}\\
 & + & \frac{1}{\exp {(\sqrt{z^{2}+\delta ^{2}y^{2}}+ry)}+1}\}\nonumber 
\end{eqnarray}
 with the following dimensionless variables \( \delta =m_{eff}/m \), \( y=mc^{2}/k_{B}T \),
\( z=\hbar kc/k_{B}T \) and \( r=\mu _{f}/mc^{2} \). The forms of the dimensionless
variables are the same as those used by Weldon in his work \cite{hw:1982},
\cite{hw:1983}. The fermion chemical potential \( \mu _{f} \) is given by
the relation (\ref{eq10}). Mutual relations between \( y \) and temperature
as well as between \( r \) and \( \mu _{f} \) result in dependence of functions
\( \Theta  \), \( \Phi  \) and \( \chi  \) on the dimensionless Fermi momentum
\( x \) and the temperature \( T \).\\
 When the temperature equals zero the Shapiro for result free nucleons can be
reproduced \cite{st:1983}
\begin{equation}
\Phi (x,0)=\frac{1}{8\pi ^{2}}\{x\sqrt{1+x^{2}}(2x^{2/3}-1)+\ln (x+\sqrt{1+x^{2}})\}
\end{equation}
 which in the nonrelativistic limit \( x<<1 \) yields 
\begin{equation}
\Phi (x,0)\rightarrow \frac{1}{15\pi ^{2}}x^{5}
\end{equation}
 and the equation of state takes the form of the polytrope \( \Gamma =5/3 \).
The forms of the functions \( \Theta  \), \( \chi  \) and \( \Phi  \) indicate
their relations with the functions \( H_{n}(r,y) \) and \( G_{n}(r,y) \) which
are used in order to evaluate thermodynamic properties of the matter \cite{ka:1989}
\begin{eqnarray}
H_{n}(r,y) & = & \int ^{\infty }_{0}\frac{z^{n-1}dz}{\sqrt{z^{2}+\delta ^{2}y^{2}}}\{\frac{1}{\exp (\sqrt{z^{2}+\delta ^{2}y^{2}}-ry)+1}\label{egg1} \\
 & + & \frac{1}{\exp (\sqrt{z^{2}+\delta ^{2}y^{2}}+ry)+1}\}\nonumber 
\end{eqnarray}
\begin{eqnarray}
G_{n}(r,y) & = & \int ^{\infty }_{0}z^{n-1}dz\{\frac{1}{\exp (\sqrt{z^{2}+\delta ^{2}y^{2}}-ry)+1}\label{egg2} \\
 & - & \frac{1}{\exp (\sqrt{z^{2}+\delta ^{2}y^{2}}+ry)+1}\}\nonumber \label{eq2} 
\end{eqnarray}
 The case where the interaction in the Lagrange function (\ref{lag}) is neglected
(free nucleons) is equivalent to the one where \( \delta  \) in equations (\ref{egg1},\ref{egg2}).
These two terms in both equations (\ref{egg1},\ref{egg2}) correspond to the
contribution of particles and antiparticles, respectively and the functions
\( H_{n}(r,y) \) and \( G_{n}(r,y) \) can be written as 
\begin{eqnarray}
H_{n}(r,y) & = & h_{n}(r,y)+h_{n}(-r,y)\nonumber \\
G_{n}(r,y) & = & g_{n}(r,y)-g_{n}(-r,y)
\end{eqnarray}
 Both the pressure \( P_{f} \) and the energy density \( \varepsilon _{f} \)
of the fermion system can be expressed with the use of the functions presented
above 
\begin{equation}
\label{eq11}
P_{f}=\frac{1}{3\pi ^{2}}P_{0}(\frac{\lambda }{\lambda _{T}})^{4}\{H_{5}(r_{p},y_{p})+H_{5}(r_{n},y_{n})+H_{5}(r_{e},y_{e})\}
\end{equation}
 where \( P_{0}=(m_{f}c^{2})/\lambda ^{3} \), the fermion Compton wavelength
\( \lambda =\hbar /(m_{f}c) \) and \( \lambda _{T}=(c\hbar )/(k_{B}T) \)
\begin{eqnarray}
\varepsilon _{f} & = & \frac{1}{\pi ^{2}}P_{0}(\frac{\lambda }{\lambda _{T}})^{4}\{H_{5}(r_{p},y_{p})+H_{5}(r_{n},y_{n})+H_{5}(r_{e},y_{e})\\
 & + & H_{3}(r_{n},y_{n})+H_{3}(r_{p},y_{p})+H_{3}(r_{e},y_{e})\}\nonumber \label{eq12} 
\end{eqnarray}
 Our first step is to find the pressure and energy density for nucleons and
muons (\( \delta =1 \)) which corresponds to the nonrelativistic limit of the
functions \( G_{n}(r,y) \) and \( H_{n}(r,y) \). Such a limit means either
the case of large mass or low temperature. Introducing the new variable \( \omega =\exp {(y-\sqrt{x^{2}+y^{2}})} \)
the following forms of the functions are achieved 
\begin{equation}
h_{5}(r,y)=\int ^{1}_{0}\frac{(-\ln {\omega })^{3/2}(2y\delta -\ln {\omega })^{3/2}d\omega }{e^{y(\delta -r)}+\omega }
\end{equation}
\begin{equation}
g_{5}(r,y)=\int ^{1}_{0}\frac{(-\ln {\omega })^{3/2}(2y\delta -\ln {\omega })^{3/2}(y\delta -\ln {\omega })d\omega }{e^{y(\delta -r)}+\omega }
\end{equation}
 The calculation of these integrals has been performed on the basis of the method
presented by Weldon in his work \cite{hw:1982}, \cite{hw:1983}. Making use
of the fact that \( |\ln {\frac{\omega }{2y\delta }}|<1 \) in this approximation
and expanding the numerators under that assumption, the functions \( h_{5} \)
and \( g_{5} \) can be written in the form of 
\begin{equation}
h_{5}(r,y)=(2y\delta )^{3/2}\sum _{k=0}^{\infty }\frac{\Gamma (\frac{5}{2})}{(2\delta y)^{k}\Gamma (\frac{3}{2}-k)k!}\int ^{1}_{0}\frac{(-\ln {\omega })^{3/2+k}d\omega }{e^{y(\delta -r)}+\omega }
\end{equation}
\begin{equation}
g_{5}(r,y)=(2y\delta )^{3/2}\sum _{k=0}^{\infty }\frac{\Gamma (\frac{5}{2})}{(2y\delta )^{k}\Gamma (\frac{3}{2}-k)k!}\int ^{1}_{0}\frac{(-\ln {\omega })^{3/2}(2y\delta -\ln {\omega })^{3/2}d\omega }{e^{y(\delta -r)}+\omega }
\end{equation}
 Having integrated the obtained equations term by term the final forms of the
function emerge: 
\begin{equation}
h_{5}(r,y)=(2y\delta )^{3/2}\Gamma (\frac{5}{2})\sum _{k=0}^{\infty }\frac{\Gamma (\frac{5}{2}+k)}{\Gamma (\frac{5}{2}-k)k!}(\frac{1}{2y\delta })^{k}Li_{k+5/2}(-e^{y(r-\delta )})
\end{equation}
\begin{eqnarray}
g_{5}(r,y) & = & (2y\delta )^{3/2}\Gamma (\frac{5}{2})\sum _{k=0}^{\infty }\frac{1}{\Gamma (\frac{5}{2}-k)k!}(\frac{1}{2y\delta })^{k}\\
 &  & (y\delta Li_{k+5/2}(-e^{y(r-\delta )})\Gamma (\frac{5}{2}+k)+\Gamma (\frac{7}{2}+k)Li_{7/2+k}(-e^{y(r-\delta )}))\nonumber 
\end{eqnarray}
 In order to calculate the contribution of electrons to the total pressure \( P_{f} \)
and the energy density \( \varepsilon _{f} \), the relativistic case should
be considered. In this very case the variable \( y \) tends to zero. It is
necessary to calculate the mentioned above functions \( H_{n}(r,y) \) and \( G_{n}(r,y) \)
in the relativistic limit, where again the method used by Weldon in \cite{hw:1982},
\cite{hw:1983} has been involved. These functions obey the recursion relations
\begin{eqnarray}
\frac{dG_{n+1}}{dy} & = & lrH_{n+1}-\frac{y}{n}G_{n-1}+\frac{y^{2}r}{n}H_{n-1}\nonumber \\
\frac{dH_{n+1}}{dy} & = & \frac{r}{y}G_{n-1}-\frac{y}{n}H_{n-1}
\end{eqnarray}
 Thus the functions \( H_{1}(r,y) \) and \( G_{1}(r,y) \) together with the
initial conditions 
\begin{eqnarray}
G_{n}(0,0) & = & 0\nonumber \\
H_{n}(0,0) & = & 2(1-2^{2-n})\Gamma (n-1)\zeta (n-1)\label{eq4} 
\end{eqnarray}
 are sufficient to determine the functions \( H_{5}(r,y) \) and \( H_{3}(r,y) \)
which in turn are indispensable to calculate the pressure and energy density.
The identity 
\begin{equation}
\frac{1}{\exp {y}+1}=\frac{1}{2}-2\sum _{0}^{\infty }{\frac{y}{y^{2}+\pi ^{2}(2n+1)^{2}}}
\end{equation}
 originated from Dolan and Jackiw \cite{dj:1974} gives us the possibility to
write the functions \( H_{1} \) and \( G_{1} \) as 
\begin{eqnarray}
H_{1}(r,y) & = & -\int _{0}^{\infty }{\frac{dz}{\sqrt{(z^{2}+y^{2})}}}\\
 & - & 4\sum _{n=0}^{\infty }\int _{0}^{\infty }{\frac{[z^{2}+y^{2}(1-r^{2})+(2n+1)^{2}\pi ^{2}]dz}{[z^{2}+y^{2}(1-r^{2})+(2n+1)^{2}\pi ^{2}]^{2}+4\pi ^{2}r^{2}y^{2}(2n+1)^{2}}}\nonumber 
\end{eqnarray}
\begin{equation}
G_{1}(r,y)=-4ry\sum _{n=0}^{\infty }\int _{0}^{\infty }{\frac{[z^{2}+y^{2}(1-r^{2})-(2n+1)^{2}\pi ^{2}]dz}{[z^{2}+y^{2}(1-r^{2})+(2n+1)^{2}\pi ^{2}]^{2}+4\pi ^{2}r^{2}y^{2}(2n+1)^{2}}}
\end{equation}
 The integrands in these equations are multiplied by the convergent factor \( z^{-\varepsilon } \)
and after its expansion as a power series in \( y \) the term by term integration
is performed. In the next step the summation over \( n \) is carried out. In
the last stage the limit \( \varepsilon \rightarrow 0 \) enables us to obtain
the final result which in the first approximation achieves the form of 
\begin{equation}
H_{1}(r,y)=-(\gamma +\ln (\frac{y}{\pi }))
\end{equation}
\begin{equation}
G_{1}(r,y)=ry
\end{equation}
 Knowing the form of functions \( H_{1}(r,y) \) and \( G_{1}(r,y) \) and the
initial conditions (\ref{eq4}) it is possible to calculate the functions \( H_{3}(r,y) \),
\( H_{5}(r,y) \), \( G_{3}(r,y) \)
\begin{equation}
H_{3}(r,y)=\frac{1}{8}(2r^{2}+2\gamma -1)y^{2}+\frac{1}{4}y^{2}\ln ({y/\pi })
\end{equation}
\begin{equation}
H_{5}(r,y)=\frac{1}{768}(8r^{4}-24r^{2}-12\gamma +9)y^{4}-\frac{1}{64}y^{4}\ln {y/\pi }
\end{equation}
\begin{equation}
G_{3}(r,y)=\frac{1}{12}r(2r^{2}-3)y^{3}
\end{equation}
 These functions are employed to express the electron pressure and energy density
according to the equations (\ref{eq11}) and (\ref{eq12}). Consequently the
relation between the total pressure \( P_{f} \) being the sum of the leptons
and nucleons pressures and the dimensionless Fermi momentum \( x \) is presented
in Fig.2. The curves in Fig, 2 are parameterized by the temperature and \( \delta  \).
The curves for the zero temperature limit are obtained for two cases with and
without (\( v=0 \)) the presence of interaction. The curve for the temperature
\( T=50\, MeV \) is the case with interactions. 
\begin{figure}
{\par\centering \resizebox*{15cm}{!}{\includegraphics{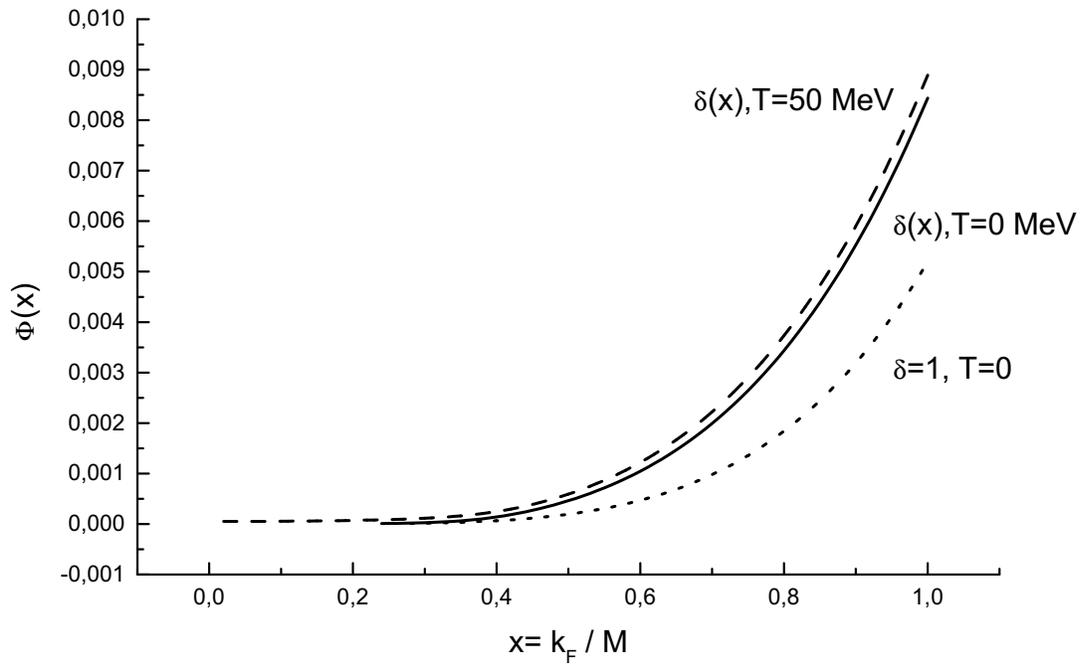}} \par}

\caption{The pressure \protect\( \Phi (x)\protect \) as the function of the Fermi momentum
\protect\( x=p_{F}/mc\protect \) }
\end{figure}

Analogous description can be made analyzing the connection between the energy
density \( \chi  \) and the parameter \( x \). See Fig.3. 
\begin{figure}
{\par\centering \resizebox*{15cm}{!}{\includegraphics{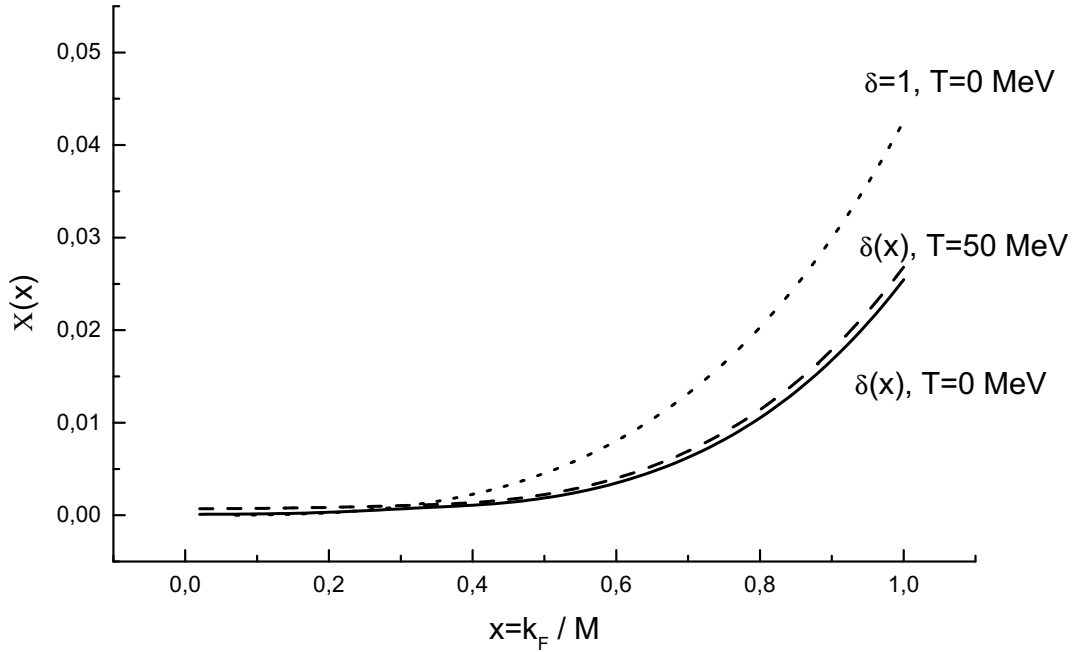}} \par}

\caption{The energy density \protect\( \chi (x)\protect \) as the function of the Fermi
momentum \protect\( x=p_{F}/mc\protect \) }
\end{figure}

\section{The neutron star}

The most important factor determining the structure of a neutron star is the
equation of state. This very equation makes possible to describe a static spherical
star solving the OTV equation. 
\begin{equation}
\frac{dm(r)}{dr}=4\pi r^{2}\rho (r)
\end{equation}
\begin{equation}
\frac{dP(r)}{dr}=-\frac{G}{r^{2}}(\rho (r)+\frac{P(r)}{c^{2}})\frac{(m(r)+\frac{4\pi }{c^{2}}P(r)r^{3})}{(1-\frac{2Gm(r)}{c^{2}r})}
\end{equation}
 Having solved this equation the pressure \( P(r) \), mass \( m(r) \) and
density \( \rho (r) \) were obtained. To achieve the total radius \( R \)
of the star the fulfillment of the condition \( P(R)=0 \) is necessary which
allows to determine the total gravitational mass of the star \( M(R) \).\\
 Introduction of the dimensionless variable \( \xi  \) which is connected with
the variable \( r \) by the relation \( r=a\xi  \) (\( a=1 \) km) enables
us to define the functions \( P(r) \), \( \rho (r) \) and \( m(r) \) in the
following form 
\begin{equation}
\rho (r)=\rho _{c}f^{\frac{3}{2}}(\xi )
\end{equation}
\begin{equation}
P(r)=P_{c}u(\xi )
\end{equation}
\begin{equation}
m(r)=M_{0}v(\xi )
\end{equation}
 Some more parameters, namely 
\begin{equation}
\lambda =\frac{GM_{o}\rho _{c}}{P_{c}a}
\end{equation}
\begin{equation}
\omega =\frac{P_{c}}{c^{2}\rho _{c}}
\end{equation}
 and 
\begin{equation}
\tau =3\frac{M_{c}}{M_{o}},\hspace {0.5cm}M_{c}=\frac{4}{3}\pi \rho _{c}a^{3}
\end{equation}
 are also needed to achieve the useful form of the OTV equation 
\begin{eqnarray}
\frac{du}{d\xi } & = & -\lambda (f(\xi )^{\frac{3}{2}}+\omega u(\xi ))\frac{v(\xi )+\omega \tau u(\xi )\xi ^{3}}{\xi ^{2}(1-\frac{r_{g}}{a}\frac{v(\xi )}{\xi })}\\
\frac{dv}{d\xi } & = & \tau f(\xi )^{\frac{3}{2}}\xi ^{2}
\end{eqnarray}
 with 
\begin{equation}
r_{g}=\frac{2GM_{o}}{c^{2}}
\end{equation}
 These equations can be solved specifying the central neutron energy density
being the energy density for \( r=0 \).\\
 The equation of state is the function of the temperature and the neutron chemical
potential which changes with the radius \( r \). Therefore the changes of the
radius influence the parameters of the star. Using the results obtained in previous
chapter it is now possible to write the functions \( P(r) \) and \( \rho (r) \)
in the form 
\begin{equation}
\rho (\mu ,T)=\rho _{0}\chi (x,T)=\rho _{c}f^{\frac{3}{2}}(\xi )
\end{equation}
\begin{equation}
P_{0}\Phi (x,T)=P_{c}u(\xi )
\end{equation}
 and the variable \( x \) can be obtained 
\begin{equation}
x=\chi ^{-1}(\frac{\rho _{c}}{\rho _{0}}f^{\frac{3}{2}}(\xi ))
\end{equation}
\begin{equation}
P_{c}u(\xi )=P_{0}\Phi (\chi ^{-1}(\frac{\rho _{c}}{\rho _{0}}f^{\frac{3}{2}}),T)
\end{equation}
 The function \( u(\xi ) \) takes the form 
\begin{equation}
u(\xi )=\frac{P_{0}}{P_{c}}\Phi (\chi ^{-1}(\frac{\rho _{c}}{\rho _{0}}f^{\frac{3}{2}}(\xi )),T).
\end{equation}
 The solution of the Oppenheimer-Volkoff-Tolman equation depicts the mass-radius
relation. Fig.4 allows us to compare the zero-temperature mass versus radius
relation with the other temperature cases with and without (\( v=0 \)) interaction
. 
\begin{figure}
{\par\centering \resizebox*{15cm}{!}{\includegraphics{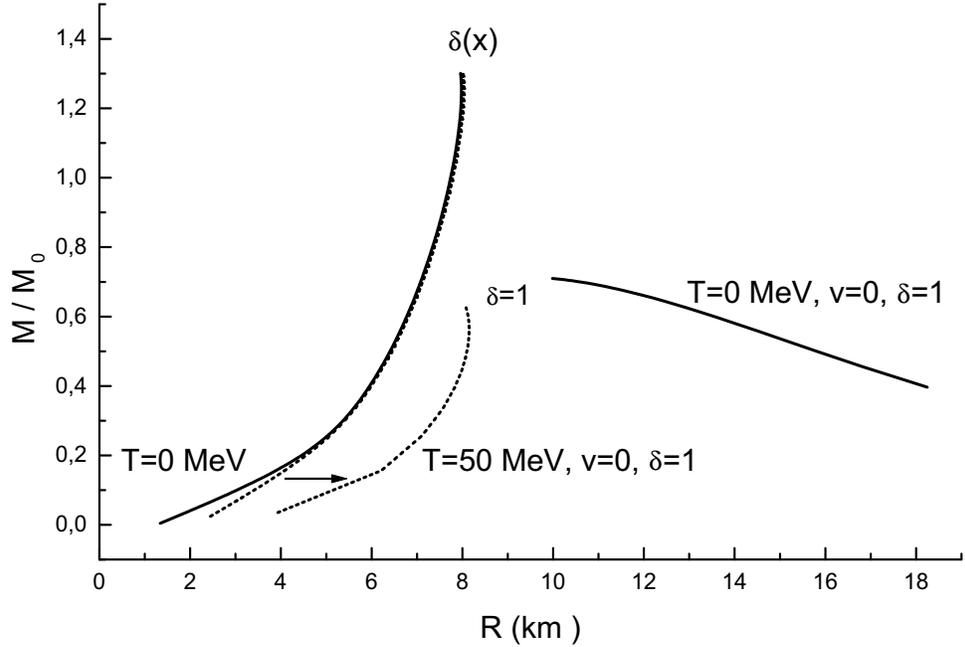}} \par}

\caption{The R-M diagram for the neutron star with interaction and without (\protect\( v=0\protect \)). }
\end{figure}

Fig.5. shows the changes of the radius as a function of the neutron density
\( \rho _{c} \) in the center of the star. 
\begin{figure}
{\par\centering \resizebox*{15cm}{!}{\includegraphics{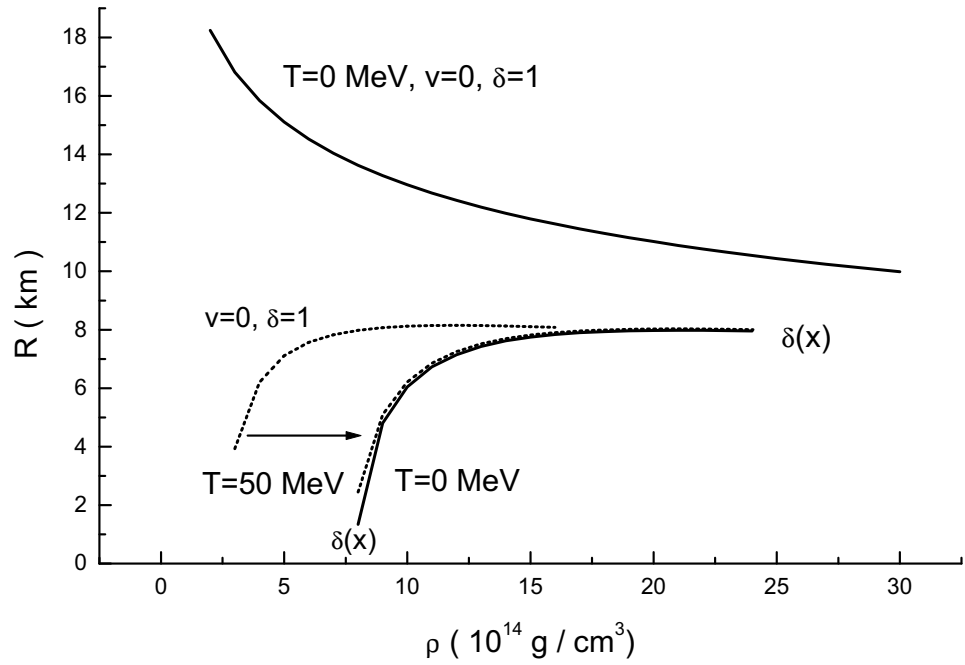}} \par}

\caption{The neutron star radius \protect\( R(km)\protect \) as a function of the neutron
density \protect\( \rho _{c}\protect \) in the center of the star.}
\end{figure}

\section{Conclusions}

Herein presented solutions of the Oppenheimer-Volkoff-Tolman equations of hydrostatic
equilibrium concern hot neutron stars. The neutron star matter in this model
consists of neutrons, protons and leptons (electrons,muons) being in \( \beta  \)-
equilibrium under the assumption that the temperature is different from zero.
It varies from 0 to 50 MeV. The objective of our work was to study the influence
of the temperature on the main parameters of a neutron star. In order to achieve
the proper form of the equation of state, which is determined only by the neutron
Fermi momentum, it is necessary to calculate either the low temperature or the
high temperature expansion of the integrals (\ref{eq1}) and (\ref{eq2}). This
very simple model presents a few global properties of the hot neutron star such
as the mass and the size.The parameters of neutron stars obtained in this simplified
model considering zero temperature and cases of finite temperatures vary from
one another. In each of the mentioned above cases stars with the same energy
density inside are considered. However, their baryon numbers are different which
makes each of them a different star with specific baryon number. In this situation
we do not deal with the thermal evolution of one star with conserved baryon
number but several separate cases. The star whose parameters at the temperature
of 50 MeV are as follows: for \( \delta =1 \) \( M=0.62\, M_{\odot } \) \( R=8.08\, km \),
for \( \delta \neq 1 \) \( M=1.3\, M_{\odot } \), \( R=8.00\, km \) At zero
temperature limit the star is characterized by the mass \( 0.7\, M_{\odot } \)
and the radius \( 10.54\, km \) for \( \delta =1 \) and the mass \( 1.24\, M_{\odot } \)
and radius \( 7.98\, km \) for \( \delta \neq 1 \). It is obvious that some
more extended models e.g. those including boson fields should be examined. Therefore
we would like to continue the subject in our next papers

.

\end{document}